\documentclass[aps,preprint]{revtex4-1}%
\usepackage{bm}
\usepackage{amsfonts}
\usepackage{amsmath}
\usepackage{amssymb}
\usepackage{graphicx}%
\begin{document}

\title{Collective modes of the order parameter in a triplet superfluid neutron liquid}
\author{L. B. Leinson}

\begin{abstract}
The complete spectrum of collective modes of the triplet order parameter in
the superfluid neutron matter is examined in the BCS approximation below the
pair-breaking threshold. The dispersion equations both for the unitary and
nonunitary excitations are derived and solved in the limit of $q\rightarrow0$
by taking into account the anisotropy of the energy gap for the case of
$P$-wave pairing. By our analysis, there is only one Goldstone mode which is
associated with the broken gauge symmetry. We found no additional Goldstone
modes associated with the broken rotational symmetry but found that the
oscillations of the total angular momentum are qualitatively similar to the
"normal-flapping" mode in the A-phase of superfluid Helium. There are also two
collective modes associated with internal vibrations of the structure of the
order parameter oscillating with $\omega\left(  T=0\right)  =1.20\Delta_{0}$
and $\omega\left(  T=0\right)  =0.61\Delta_{0}$.

\end{abstract}
\maketitle

\affiliation{Institute of Terrestrial Magnetism, Ionosphere and Radio Wave Propagation RAS (IZMIRAN),
142190 Troitsk, Moscow Region, Russia}

\startpage{1}

\section{Introduction}

The neutron star evolution is governed mostly by microscopic processes
occuring in its volume and depend substantially on the spectrum of thermal
excitations that can exist in the bulk baryon matter. Strong and
electromagnetic interactions form numerous collective excited states in the
normal (nonsuperfluid) component of the neutron star matter which have been
extensively studied by many authors. \cite{H78}-\cite{PB}.

The purpose of this paper is to present the complete spectrum of collective
modes in the superfluid phase of superdense neutron matter. It is generally
accepted that the pair condensation in superdense neutron matter occurs into
the $^{3}P_{2}$ state (with a small admixture of $^{3}F_{2}$ ). Such a model
is based on the properties of the bare $NN$ interaction, which contains a
relatively strong spin-orbit interaction in this channel. Strictly speaking,
it is not really clear that a $^{3}P_{2}$ pairing instability will survive at
the relevant densities in neutron stars (double nuclear saturation density, or
more), due to uncertainties in the 3$N$ interaction. Nevertheless this model
is conventionally used for estimates of neutrino energy losses in the minimal
cooling scenarios of neutron stars.

We do not consider the well discussed processes of pair breaking and formation
\cite{YKL}, \cite{L10a} but focus on the undamped collective oscillations of
the order parameter. The collective excitations below the pair-breaking
threshold can play an important role in the thermodynamic properties of the
neutron liquid at lowest temperatures when the other possible excitations are
strongly suppressed by the superfluidity. Some properties of the anisotropic
Goldstone mode in the neutron condensate were discussed in Ref. \cite{L11a},
and the existence of two spin-wave modes was predicted recently \cite{L10b}%
-\cite{L11c} in the average-angle approximation. In the present paper we
consider the solution to the dispersion equations by taking into account the
anisotropy of the energy gap and find the whole spectrum of the undamped
collective oscillations of the $^{3}P_{2}$ order parameter.

In applications to neutron stars it is customary \cite{hof} to consider the
$^{3}P_{2}$ state with a preferred magnetic quantum number $M=0$.
Sophisticated calculations \cite{Khodel}-\cite{0203046} have shown that,
besides the above one-component state, there are also multicomponent states
involving several magnetic quantum numbers that compete in energy and
represent various phase states of the condensate dependent on the temperature.
A delicate difference in the above gap magnitudes occurs owing to small tensor
interactions between paired quasiparticles. If we neglect the tensor
interactions the multicomponent states are degenerate and can be obtained from
the state $M=0$ by trivial rotation of the frame in space. We therefore will
focus on the $^{3}P_{2}$ neutron pairing into a state with $M=0$.

Previously eigen-modes of the order parameter have been thoroughly studied in
the superfluid liquid $^{3}$He \cite{Wolfe73}-\cite{Wolfe1}. The complete
analysis of the order-parameter collective modes in anisotropic $^{3}$He-A is
given in Ref. \cite{W}. The pairing interaction in $^{3}$He is invariant with
respect to the rotation of spin and orbital coordinates separately. This
admits spin fluctuations independent of the orbital coordinates. In contrast,
the spin-triplet neutron condensate arises in the high-density neutron matter
mostly owing to the attractive spin-orbit interactions which do not possess
the above symmetry. Therefore the results obtained for liquid $^{3}$He can not
be applied directly to the spin-triplet neutron superfluid, where the most
attractive channel of interactions corresponds to spin, orbital, and total
angular momenta $S=1$, $L=1$, and $J=2$, respectively, and pairs
quasiparticles into the $^{3}P_{2}$ states with the projection of the total
angular momentum $M=0,\pm1,\pm2$.

As is well known the strong interactions in the neutron matter are not
restricted by the pairing forces. One can expect that the Fermi-liquid effects
are able to renormalize the collective modes. We hope however that the
principal effect occurs in the case of the Goldstone mode, where the sound
velocity is strongly renormalized by the particle-hole interactions. This
effect was already considered in Ref. \cite{L11a}. Since the Landau parameters
are unknown for a dense asymmetric baryon matter it would be meaningless to
complicate the calculations by the Fermi liquid effects. Therefore we discard
residual particle-hole interactions and consider the problem in the BCS model.

The paper is organized as follows. Section II contains some preliminary notes.
We consider the nonequilibrium gap equations for the case of spin-orbit
pairing forces and discuss the renormalizations which transform the standard
gap equations to a very simple form valid near the Fermi surface. In Sec. III
we consider small deviations of the condensate from equilibrium caused by weak
external fields and derive the equations for the relevant anomalous
three-point vertices. In Sec. IV we derive, in the BCS approximation, the
dispersion equations for eigenmodes of the order parameter and find the whole
spectrum of the undamped oscillations. Section V represents a discussion of
the obtained results and the conclusion. Some intermediate and additional
calculations are contained in three appendices for a more deep understanding.

Throughout this paper, we use the standard model of weak interactions, the
system of units $\hbar=c=1$ and the Boltzmann constant $k_{B}=1$.

\section{Formalism}

We employ the Matsubara calculation technique for the system in thermal
equilibrium and use the standard notation for ordinary propagators of a
quasiparticle and a hole, $\hat{G}\left(  p_{\kappa}.\mathbf{p}\right)
=\parbox{1cm}{\includegraphics[width=1cm]{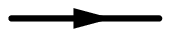}}$, $\hat{G}^{-}\left(
p_{\kappa}.\mathbf{p}\right)  \equiv\hat{G}\left(  -p_{\kappa},-\mathbf{p}%
\right)  =\parbox{1cm}{\includegraphics[width=1cm,angle=180]{Gn.eps}}$, and
for anomalous propagators $\hat{F}^{\left(  1\right)  }%
=\parbox{1cm}{\includegraphics[width=1cm]{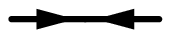}}$\thinspace, $\hat
{F}^{\left(  2\right)  }=\parbox{1cm}{\includegraphics[width=1cm]{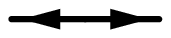}}$ in
the momentum representation, where $\mathbf{p}$ is the quasiparticle momentum,
and $p_{\kappa}=\left(  2\kappa+1\right)  \pi T$ with $\kappa=0,\pm1,\pm2,...$
is the fermionic Matsubara frequency which depends on the temperature $T$.

The triplet order parameter, $\hat{D}\equiv D_{\alpha\beta}\left(
\mathbf{n}\right)  $, in the neutron superfluid represents a symmetric matrix
in spin space $\left(  \alpha,\beta=\uparrow,\downarrow\right)  $ which can be
written as $\hat{D}\left(  \mathbf{n}\right)  =\Delta\mathbf{\bar{b}%
}\bm{\hat{\sigma}}\hat{g}$, where $\bm{\hat{\sigma}}=\left(  \hat{\sigma}%
_{1},\hat{\sigma}_{2},\hat{\sigma}_{3}\right)  $ are Pauli spin matrices, and
$\hat{g}=i\hat{\sigma}_{2}$. The angular dependence of the order parameter is
represented by Cartesian components of the unit vector $\mathbf{n=p}/p$ which
involves the polar angles on the Fermi surface,
\begin{equation}
n_{1}=\sin\theta\ \cos\varphi,\ \ \ n_{2}=\sin\theta\ \sin\varphi
,\ \ \ n_{3}=\cos\theta. \label{n}%
\end{equation}
The temperature-dependent gap amplitude $\Delta\left(  T\right)  $ is a real
constant (on the Fermi surface), and $\mathbf{\bar{b}}\left(  \mathbf{n}%
\right)  $ is a real vector in spin space which we normalize by the condition
\begin{equation}
\left\langle \bar{b}^{2}\left(  \mathbf{n}\right)  \right\rangle =1~.
\label{Norm}%
\end{equation}
Hereafter we use the angle brackets to denote angle averages,
\begin{equation}
\left\langle ...\right\rangle \equiv\frac{1}{4\pi}\int d\mathbf{n}\cdot
\cdot\cdot=\frac{1}{2}\int_{-1}^{1}dn_{3}\int_{0}^{2\pi}\frac{d\varphi}{2\pi
}\cdot\cdot\cdot. \label{av}%
\end{equation}

The analytic form of the quasiparticle propagators can be written as
\begin{equation}
\hat{G}\left(  p_{\kappa},\mathbf{p}\right)  =G\left(  p_{\kappa}%
,\mathbf{p}\right)  \hat{1},~\ \ \hat{F}^{\left(  1\right)  }=F\left(
p_{\kappa},\mathbf{p}\right)  \mathbf{\bar{b}}\hat{\bm{\sigma}}\hat{g},
\label{eqpr}%
\end{equation}%
\begin{equation}
\ \hat{F}^{\left(  2\right)  }=\hat{F}^{+}\left(  -p_{\kappa},-\mathbf{p}%
\right)  =\hat{g}\hat{\bm{\sigma}}\mathbf{\bar{b}}F\left(  p_{\kappa
},\mathbf{p}\right)  , \label{F2}%
\end{equation}
where we define the scalar Green functions%
\begin{equation}
G\left(  p_{\kappa},\mathbf{p}\right)  =\frac{-ip_{\kappa}-\varepsilon_{p}%
}{p_{\kappa}^{2}+E_{\mathbf{p}}^{2}},~\ \ F\left(  p_{\kappa},\mathbf{p}%
\right)  =\frac{\Delta}{p_{\kappa}^{2}+E_{\mathbf{p}}^{2}}. \label{GF}%
\end{equation}
The quasiparticle energy is given by%
\begin{equation}
E_{\mathbf{p}}=\sqrt{\varepsilon_{p}^{2}+\Delta^{2}\bar{b}^{2}\left(
\mathbf{n}\right)  } \label{Ep}%
\end{equation}
with
\begin{equation}
\varepsilon_{p}\simeq\upsilon_{F}\left(  p-p_{F}\right)  , \label{ksi}%
\end{equation}
where $\upsilon_{F}\ll1$ is the Fermi velocity of the nonrelativistic neutrons.

At equilibrium, the gap matrix $\hat{D}$ is connected to the anomalous Green's
function $\hat{F}$ by the gap equation \cite{gapEq}
\begin{equation}
D_{\alpha\beta}\left(  \mathbf{p}\right)  =-T\sum_{\kappa}\int\frac
{d^{3}p^{\prime}}{8\pi^{3}}\Gamma_{\alpha\gamma,\beta\delta}\left(
\mathbf{p,p}^{\prime}\right)  F_{\gamma\delta}\left(  p_{\kappa}%
.\mathbf{p}^{\prime}\right)  , \label{DGeq}%
\end{equation}
where $\Gamma_{\alpha\beta,\gamma\delta}\left(  \mathbf{p,p}^{\prime}\right)
$ stands for the block of the interaction diagrams irreducible in the channel
of two quasiparticles.

We consider small periodic departure from equilibrium of the form $\propto
\exp\left(  i\mathbf{qr}-i\omega t\right)  $. By assuming that the
interactions between quasiparticles are essentially instantaneous on the time
scale of the oscillation frequencies the nonequilibrium (time-dependent) gap
matrices $\mathfrak{D}_{\alpha\beta}\left(  \omega\mathbf{,q;p}\right)
\exp\left(  i\mathbf{qr}-i\omega t\right)  $ must obey the self-consistency
conditions (gap equations) which depend now on the energy $\omega$ and space
momentum $\mathbf{q}$ of the perturbation. Then the order parameters
$\mathfrak{D}_{\alpha\beta}^{\left(  1,2\right)  }\left(  \omega
\mathbf{,q;p}\right)  $ are to be found as the analytic continuation of the
functions $\mathfrak{D}_{\alpha\beta}^{\left(  1,2\right)  }\left(
\omega_{\eta}\mathbf{,q;p}\right)  $ connected to the anomalous Green
functions $\mathcal{\hat{F}}_{\gamma\delta}^{\left(  1,2\right)  }\left(
\omega_{\eta}\mathbf{,q};p_{\kappa}\mathbf{,p}\right)  $ by the equations%
\begin{equation}
\mathfrak{D}_{\alpha\beta}^{\left(  1\right)  }\left(  \omega_{\eta
}\mathbf{,q;p}\right)  =-T\sum_{\kappa}\int\frac{d^{3}p^{\prime}}{8\pi^{3}%
}\Gamma_{\alpha\gamma,\beta\delta}\left(  \mathbf{p,p}^{\prime}\right)
\mathcal{\hat{F}}_{\gamma\delta}^{\left(  1\right)  }\left(  \omega_{\eta
}\mathbf{,q;}p_{\kappa},\mathbf{p}^{\prime}\right)  , \label{Deq}%
\end{equation}%
\begin{equation}
\mathfrak{D}_{\alpha\beta}^{\left(  2\right)  }\left(  \omega_{\eta
}\mathbf{,q;p}\right)  =-T\sum_{\kappa}\int\frac{d^{3}p^{\prime}}{8\pi^{3}%
}\Gamma_{\beta\delta,\alpha\gamma}\left(  \mathbf{p}^{\prime}\mathbf{,p}%
\right)  \mathcal{\hat{F}}_{\gamma\delta}^{\left(  2\right)  }\left(
\omega_{\eta}\mathbf{,q;}p_{\kappa},\mathbf{p}^{\prime}\right)  , \label{Deq2}%
\end{equation}
where $\omega_{\eta}=2i\pi\eta T$ with $\eta=0,\pm1,\pm2,...$ is the bosonic
Matsubara frequency.

In these equations, the integration goes over infinite momentum space while
the quasiparticle approach is valid only near the Fermi surface. To get rid of
the integration over the regions far from the Fermi surface in Eqs.
(\ref{Deq}) and (\ref{Deq2}), we renormalize the interaction as suggested in
Refs. \cite{Larkin}, \cite{Leggett}. We define
\begin{align}
\Gamma_{\alpha\gamma,\beta\delta}^{r}\left(  \mathbf{p,p}^{\prime};T\right)
&  =\Gamma_{\alpha\gamma,\beta\delta}\left(  \mathbf{p,p}^{\prime}\right)
\nonumber\\
&  -T\sum_{\kappa}\int\frac{dp^{\prime\prime}p^{\prime\prime2}}{\pi^{2}%
\varrho}\Gamma_{\alpha\mu,\beta\nu}\left(  \mathbf{p,p}^{\prime\prime}\right)
\left(  GG^{-}\right)  _{N}^{\prime\prime}\Gamma_{\mu\gamma,\nu\delta}%
^{r}\left(  \mathbf{p}^{\prime\prime}\mathbf{,p}^{\prime};T\right)  ,
\label{Gr}%
\end{align}
where $\varrho=p_{F}M^{\ast}/\pi^{2}$, the effective mass of a neutron
quasiparticle is defined as $M^{\ast}=p_{F}/\upsilon_{F}$, and the product
$\left(  GG^{-}\right)  _{N}^{\prime\prime}\equiv G_{N}\left(  p_{\kappa
},\mathbf{p}^{\prime\prime}\right)  G_{N}\left(  -p_{\kappa},-\mathbf{p}%
^{\prime\prime}\right)  $ is evaluated in the normal (nonsuperfluid) state by
assuming $G_{N}\left(  p_{\kappa},\mathbf{p}\right)  =\left(  ip_{\kappa
}-\varepsilon_{p}\right)  ^{-1}$.

By acting from the left onto the both sides of Eqs. (\ref{Deq}) and
(\ref{Deq2}) by the operator
\begin{equation}
1-T\sum_{\kappa}\int\frac{dp^{\prime\prime}p^{\prime\prime2}}{\pi^{2}\varrho
}\Gamma_{\alpha\mu,\beta\nu}^{r}\left(  \mathbf{p,p}^{\prime\prime};T\right)
\left(  GG^{-}\right)  _{N}^{\prime\prime}\cdot\cdot\cdot, \label{oper}%
\end{equation}
and making use of the fact that
\begin{align}
\Gamma_{\alpha\gamma,\beta\delta}^{r}\left(  \mathbf{p,p}^{\prime};T\right)
&  =\Gamma_{\alpha\gamma,\beta\delta}\left(  \mathbf{p,p}^{\prime}\right)
\nonumber\\
&  -T\sum_{\kappa}\int\frac{dp^{\prime\prime}p^{\prime\prime2}}{\pi^{2}%
\varrho}\Gamma_{\alpha\mu,\beta\nu}^{r}\left(  \mathbf{p,p}^{\prime\prime
};T\right)  \left(  GG^{-}\right)  _{N}^{\prime\prime}\Gamma_{\mu\gamma
,\nu\delta}\left(  \mathbf{p}^{\prime\prime}\mathbf{,p}^{\prime}\right)  ,
\label{GR}%
\end{align}
we can recast Eqs. (\ref{Deq}) and (\ref{Deq2}) to the form
\begin{equation}
\mathfrak{D}_{\alpha\beta}^{\left(  1\right)  }\left(  \mathbf{p}\right)
=-T\sum_{\kappa}\int\frac{d^{3}p^{\prime}}{8\pi^{3}}\Gamma_{\alpha\gamma
,\beta\delta}^{r}\left(  \mathbf{p,p}^{\prime}\right)  \left[  \mathcal{\hat
{F}}_{\gamma\delta}^{\left(  1\right)  }-\mathfrak{D}_{\gamma\delta}^{\left(
1\right)  }G_{N}G_{N}^{-}\right]  _{\mathbf{p}^{\prime}}, \label{gap}%
\end{equation}%
\begin{equation}
\mathfrak{D}_{\alpha\beta}^{\left(  2\right)  }\left(  \mathbf{p}\right)
=-T\sum_{\kappa}\int\frac{d^{3}p^{\prime}}{8\pi^{3}}\Gamma_{\beta\delta
,\alpha\gamma}^{r}\left(  \mathbf{p}^{\prime}\mathbf{,p}\right)  \left[
\mathcal{\hat{F}}_{\gamma\delta}^{\left(  2\right)  }-\mathfrak{D}%
_{\gamma\delta}^{\left(  2\right)  }G_{N}G_{N}^{-}\right]  _{\mathbf{p}%
^{\prime}}. \label{gapp}%
\end{equation}
For brevity we omit the dependence of functions on $\omega_{\eta}$ and
$\mathbf{q}$. The product $G_{N}G_{N}^{-}$ is to be evaluated for
$\omega_{\eta}=0$, $\mathbf{q}=0$\textbf{. }In these equations the integrand
decreases very rapidly with the distance from the Fermi surface. Since we are
interested in the processes near the Fermi surface, we can replace all smooth
functions of the momentum $p$ with their expressions at $p=p_{F}$.

The pairing in the high-density neutron matter is mostly caused by the
attractive spin-orbit interactions. In the vector notation one can write the
renormalized $^{3}P_{2}$ interaction in the form
\begin{equation}
\varrho\Gamma_{\alpha\gamma,\beta\delta}^{r}\left(  \mathbf{p,p}^{\prime
}\right)  =-V\sum_{M=-2}^{2}\left[  \mathbf{b}_{M}(\mathbf{n})\cdot
\hat{\bm{\sigma}}\hat{g}\right]  _{\alpha\beta}\left[  \hat{g}\hat
{\bm{\sigma}}\cdot\mathbf{b}_{M}^{\ast}(\mathbf{n}^{\prime})\right]
_{\gamma\delta}~, \label{Gam}%
\end{equation}
where $V$ is the constant amplitude taken at the Fermi surface, and
$\mathbf{b}_{M}\left(  \mathbf{n}\right)  $ are the vectors in spin space
which generate standard spin-angle matrices according to
\begin{equation}
\frac{1}{\sqrt{8\pi}}\mathbf{b}_{M}(\mathbf{n})\bm{\hat{\sigma}}\hat{g}%
\equiv\sum_{M_{S}+M_{L}=M}\left(  \frac{1}{2}\frac{1}{2}\alpha\beta
|1M_{S}\right)  \left(  11M_{S}M_{L}|2M\right)  Y_{1,M_{L}}\left(
\mathbf{n}\right)  . \label{bm}%
\end{equation}
These are given by
\begin{align}
\mathbf{b}_{0}  &  =\sqrt{1/2}\left(  -n_{1},-n_{2},2n_{3}\right)
,\mathbf{b}_{1}=-\sqrt{3/4}\left(  n_{3},in_{3},n_{1}+in_{2}\right)
,\nonumber\\
\mathbf{b}_{2}  &  =\sqrt{3/4}\left(  n_{1}+in_{2},in_{1}-n_{2},0\right)
,\mathbf{b}_{-M}=\left(  -\right)  ^{M}\mathbf{b}_{M}^{\ast}. \label{b012}%
\end{align}
The vectors are mutually orthogonal and are normalized by the condition%
\begin{equation}
\left\langle \mathbf{b}_{M^{\prime}}^{\ast}\mathbf{b}_{M}\right\rangle
=\delta_{M^{\prime}M}. \label{lmnorm}%
\end{equation}

It is convenient to divide the integration over the momentum space into
integration over the solid angle and over the energy according to%
\begin{equation}
\int\frac{d^{3}p}{\left(  2\pi\right)  ^{3}}\operatorname*{Tr}\left(
...\right)  =\varrho\int\frac{d\mathbf{n}}{4\pi}\int_{-\infty}^{\infty
}d\varepsilon_{p}\frac{1}{2}\operatorname*{Tr}\left[  ...\right]  \cdot
\cdot\cdot, \label{d3p}%
\end{equation}
and denote%
\[
\sum_{p,\kappa}...\equiv\int_{-\infty}^{\infty}d\varepsilon_{p}T\sum_{\kappa
}\cdot\cdot\cdot.
\]
Then instead of Eqs. (\ref{gap}) and (\ref{gapp}) we obtain
\begin{align}
\mathfrak{\hat{D}}^{\left(  1\right)  }\left(  \mathbf{n}\right)   &
=V\sum_{M}\mathbf{b}_{M}\hat{\left(  \mathbf{n}\right)  \bm{\sigma}}\hat
{g}\nonumber\\
&  \times\left\langle \sum_{p,\kappa}\frac{1}{2}\operatorname*{Tr}\left[
\hat{g}\left(  \hat{\bm{\sigma}}\mathbf{b}_{M}^{\ast}\right)  \left(
\mathcal{\hat{F}}^{\left(  1\right)  }-\mathfrak{\hat{D}}^{\left(  1\right)
}G_{N}G_{N}^{-}\right)  \right]  \right\rangle _{\mathbf{n}^{\prime}},
\label{eqD1}%
\end{align}%
\begin{align}
\mathfrak{\hat{D}}^{\left(  2\right)  }\left(  \mathbf{n}\right)   &
=V\sum_{M}\hat{g}\hat{\bm{\sigma}}\mathbf{b}_{M}\left(  \mathbf{n}\right)
\nonumber\\
&  \times\left\langle \sum_{p,\kappa}\frac{1}{2}\operatorname*{Tr}\left[
\left(  \mathbf{b}_{M}^{\ast}\hat{\bm{\sigma}}\right)  \hat{g}\left(
\mathcal{\hat{F}}^{\left(  2\right)  }-\mathfrak{\hat{D}}^{\left(  2\right)
}G_{N}G_{N}^{-}\right)  \right]  \right\rangle _{\mathbf{n}^{\prime}},
\label{eqD2}%
\end{align}
where the dependence of functions on $\omega_{\eta}$ and $\mathbf{q}$ is again omitted.

In the absence of external fields the equilibrium order parameters are of the
simple form%
\begin{equation}
\hat{D}^{\left(  1\right)  }\left(  \mathbf{n}\right)  =\hat{D}\left(
\mathbf{n}\right)  =\Delta\mathbf{\bar{b}}\bm{\sigma}\hat{g}, \label{Delta}%
\end{equation}%
\begin{equation}
\hat{D}^{\left(  2\right)  }\left(  \mathbf{n}\right)  =\hat{D}^{+}\left(
-\mathbf{n}\right)  =\Delta\hat{g}\hat{\bm{\sigma}}\mathbf{\bar{b}}.
\label{Delta2}%
\end{equation}
From Eqs. (\ref{eqD1}) and (\ref{eqD2}) we obtain%
\begin{equation}
\Delta\mathbf{\bar{b}}\left(  \mathbf{n}\right)  =V\sum_{M}\mathbf{b}%
_{M}\left(  \mathbf{n}\right)  \left\langle \mathbf{b}_{M}^{\ast}%
\mathbf{\bar{b}}\sum_{p,\kappa}\left(  F-\Delta G_{N}G_{N}^{-}\right)
\right\rangle _{\mathbf{n}^{\prime}}. \label{beq1}%
\end{equation}
Using also the fact that $F\equiv\Delta\left(  GG^{-}+\bar{b}^{2}FF\right)  $,
Eq. (\ref{beq1}) can be written in the form%
\begin{equation}
\mathbf{\bar{b}}\left(  \mathbf{n}\right)  =V\sum_{M}\mathbf{b}_{M}%
(\mathbf{n})\left\langle \mathbf{\bar{b}b}_{M}^{\ast}A\right\rangle ,
\label{bbareq}%
\end{equation}
where the function $A$\ is defined as%
\begin{equation}
A\left(  \mathbf{n}\right)  =\int_{-\infty}^{\infty}d\varepsilon_{p}%
T\sum_{\kappa}\left(  GG^{-}-G_{N}G_{N}^{-}+\bar{b}^{2}FF\right)  . \label{A}%
\end{equation}
Its explicit form is given in Eq. (\ref{Aex}).

With the aid of the expansion%
\begin{equation}
\mathbf{\bar{b}}\left(  \mathbf{n}\right)  =\sum_{M}\left\langle
\mathbf{\bar{b}b}_{M}^{\ast}\right\rangle \mathbf{b}_{M}(\mathbf{n}),
\label{bbar}%
\end{equation}
and of Eqs. (\ref{Norm}) and (\ref{lmnorm}) one can recast Eq. (\ref{bbareq})
as
\begin{equation}
1=V\left\langle \bar{b}^{2}A\right\rangle . \label{GAP}%
\end{equation}

\section{Vertex equations}

\begin{figure}[h]
\includegraphics{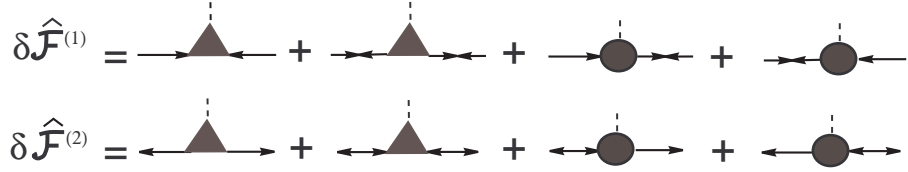}\caption{Graphs corresponding to linear corrections to
the anomalous propagators. The external field is shown by dashed lines. The
anomalous vertices are shown by filled triangles. Filled circles are dressed
ordinary vertices.}%
\label{fig1}%
\end{figure}

We now consider the order parameter of the superfluid Fermi-liquid in weak
external fields $A^{\mu}\left(  \omega,\mathbf{q}\right)  \exp\left(
i\mathbf{qr}-i\omega t\right)  $. The field interaction with a superfluid
should be described with the aid of four effective three-point vertices. There
are two ordinary effective vertices, $\hat{\tau}_{\mu}\left(  \omega
,\mathbf{q;p}\right)  $ and $\hat{\tau}_{\mu}^{-}\left(  \mathbf{p}\right)
=\hat{\tau}_{\mu}^{T}\left(  -\mathbf{p}\right)  $, corresponding to the
creation of a particle and a hole by the field (These differ by direction of
fermion lines), and two anomalous vertices $\hat{T}_{\mu}^{\left(  1,2\right)
}\left(  \omega,\mathbf{q;p}\right)  $ corresponding to the creation of two
particles or two holes. The latter describe the linear departure from
equilibrium of the matrix order parameters caused by the external field of
wave vector $\mathbf{q}$ and frequency $\omega$:
\begin{equation}
\delta\hat{D}^{\left(  1,2\right)  }\left(  \omega\mathbf{,q;p}\right)
=e\hat{T}_{\mu}^{\left(  1,2\right)  }\left(  \omega,\mathbf{q;p}\right)
A^{\mu}\left(  \omega,\mathbf{q}\right)  . \label{dD}%
\end{equation}
Here the coupling constant $e$ is specified by the external field and a
summation over repeated Dirac indices $%
\mu
=0,1,2,3$ is implied. The linearization of Eqs. (\ref{eqD1}) and (\ref{eqD2})
by taking into account Eq. (\ref{beq1}) results in equations for the anomalous
three-point vertices $\hat{T}_{\mu}^{\left(  1,2\right)  }(\omega
\mathbf{,q;n})$
\begin{align}
\hat{T}_{\mu}^{\left(  1\right)  }  &  =-V\sum_{M}\mathbf{b}_{M}%
\hat{\bm{\sigma}}\hat{g}\nonumber\\
&  \times\left\langle \sum_{p,\kappa}\frac{1}{2}\operatorname*{Tr}\left[
\hat{g}\hat{\bm{\sigma}}\mathbf{b}_{M}^{\ast}\left(  \frac{\delta
\mathcal{\hat{F}}^{\left(  1\right)  }}{eA^{\mu}}-\hat{T}_{\mu}^{\left(
1\right)  }G_{N}G_{N}^{-}\right)  \right]  \right\rangle , \label{aF1}%
\end{align}%
\begin{align}
\hat{T}_{\mu}^{\left(  2\right)  }  &  =-V\sum_{M}\hat{g}\hat{\bm{\sigma}}%
~\mathbf{b}_{M}\nonumber\\
&  \times\left\langle \sum_{p,\kappa}\frac{1}{2}\operatorname*{Tr}\left[
\mathbf{b}_{M}^{\ast}\hat{\bm{\sigma}}\hat{g}\left(  \frac{\delta
\mathcal{\hat{F}}^{\left(  2\right)  }}{eA^{\mu}}-\hat{T}_{\mu}^{\left(
2\right)  }G_{N}G_{N}^{-}\right)  \right]  \right\rangle , \label{aF2}%
\end{align}
where the product $G_{N}G_{N}^{-}$ is to be evaluated for $\omega
=0,\mathbf{q}=0$. The corrections to the anomalous propagators $\delta
\mathcal{\hat{F}}^{\left(  1,2\right)  }$ arise owing to the variation of the
order parameters and the ordinary self-energy of a quasiparticle in the
external field and can be depicted symbolically by the graphs in Fig.
\ref{fig1}.

Analytically one has%
\begin{align}
\frac{\delta\mathcal{\hat{F}}^{\left(  1\right)  }}{eA^{\mu}}  &  \simeq
\hat{G}\left(  P_{+}\right)  \hat{T}_{\mu}^{\left(  1\right)  }\hat{G}%
^{-}\left(  P_{-}\right)  +\hat{F}^{\left(  1\right)  }\left(  P_{+}\right)
\hat{T}_{\mu}^{\left(  2\right)  }\hat{F}^{\left(  1\right)  }\left(
P_{-}\right) \nonumber\\
&  +\hat{G}\left(  P_{+}\right)  \hat{\tau}_{\mu}\hat{F}^{\left(  1\right)
}\left(  P_{-}\right)  +\hat{F}^{\left(  1\right)  }\left(  P_{+}\right)
\hat{\tau}_{\mu}^{-}\hat{G}^{-}\left(  P_{-}\right)  , \label{dF1}%
\end{align}
and%
\begin{align}
\frac{\delta\mathcal{\hat{F}}^{\left(  2\right)  }}{eA^{\mu}}  &  \simeq
\hat{G}^{-}\left(  P_{+}\right)  \hat{T}_{\mu}^{\left(  2\right)  }\hat
{G}\left(  P_{-}\right)  +\hat{F}^{\left(  2\right)  }\left(  P_{+}\right)
\hat{T}_{\mu}^{\left(  1\right)  }\hat{F}^{\left(  2\right)  }\left(
P_{-}\right) \nonumber\\
&  +\hat{F}^{\left(  2\right)  }\left(  P_{+}\right)  \hat{\tau}_{\mu}\hat
{G}\left(  P_{-}\right)  +\hat{G}^{-}\left(  P_{+}\right)  \hat{\tau}_{\mu
}^{-}\hat{F}^{\left(  2\right)  }\left(  P_{-}\right)  , \label{dF2}%
\end{align}
where we denote $P_{+}=\left\{  p_{\kappa}+\omega_{\eta},\mathbf{p}+\frac
{1}{2}\mathbf{q}\right\}  $ and $P_{-}=\left\{  p_{\kappa},\mathbf{p}-\frac
{1}{2}\mathbf{q}\right\}  $.

We shall examine collective excitations of the condensate below the
pair-breaking threshold. The eigenmodes are known to manifest themselves as
sharp resonances of the linear response of the medium onto external
perturbations. Throughout this paper we consider the case of $^{3}P_{2}$
neutron pairing, when quasiparticles pair in the most attractive channel with
spin, orbital and total angular momenta, $S=1,L=1,J=2$, respectively. This
allows one to search for the anomalous vertices near the Fermi surface in the
form of expansions over the eigen functions of the total angular momentum
$(J,M)$ with $J=2$ and $M=0,\pm1,\pm2$:
\begin{equation}
\hat{T}^{\left(  1\right)  }\left(  \omega,\mathbf{q;n}\right)  =\sum_{M}%
B_{M}^{\left(  1\right)  }\left(  \omega,\mathbf{q}\right)  \left(
\bm{\hat{\sigma}}\mathbf{b}_{M}\right)  \hat{g}, \label{T1A}%
\end{equation}%
\begin{equation}
\hat{T}^{\left(  2\right)  }\left(  \omega,\mathbf{q;n}\right)  =\sum_{M}%
B_{M}^{\left(  2\right)  }\left(  \omega,\mathbf{q}\right)  \hat{g}\left(
\bm{\hat{\sigma}}\mathbf{b}_{M}\right)  . \label{T2A}%
\end{equation}
Hereafter we omit the Dirac indices by assuming that the equations are valid
for each of the vertex components. Inserting expressions (\ref{T1A}) and
(\ref{T2A}) into Eqs. (\ref{aF1}) and (\ref{aF2}) by making use of Eq.
(\ref{FFp}), one can obtain the set of equations for $B_{M}^{\left(
1,2\right)  }$. The amplitude of the pairing interaction $V$ can be eliminated
by virtue of the gap equation (\ref{GAP}). Substantial simplifications can be
further done by making use of Eqs. (\ref{Leg}), (\ref{gf}), and (\ref{ff1}).

In this way we obtain two uncoupled sets of equations,
\begin{align}
&  \sum_{M^{\prime}}2\left\langle \left[  \left(  \Omega^{2}-\eta^{2}-\bar
{b}^{2}\right)  \mathbf{b}_{M}^{\ast}\mathbf{b}_{M^{\prime}}+\left(
\mathbf{b}_{M}^{\ast}\mathbf{\bar{b}}\right)  \left(  \mathbf{b}_{M^{\prime}%
}\mathbf{\bar{b}}\right)  \right]  \mathcal{I}_{FF}\right\rangle B_{M^{\prime
}}^{-}\nonumber\\
&  +\left\langle \left(  b_{M}^{2}-\bar{b}^{2}\right)  A\right\rangle
B_{M}^{-}=R_{M}^{-}, \label{Bm}%
\end{align}
and%
\begin{align}
&  \sum_{M^{\prime}}~2\left\langle \left[  \left(  \Omega^{2}-\eta^{2}\right)
\mathbf{b}_{M}^{\ast}\mathbf{b}_{M^{\prime}}-\left(  \mathbf{b}_{M}^{\ast
}\mathbf{\bar{b}}\right)  \left(  \mathbf{b}_{M^{\prime}}\mathbf{\bar{b}%
}\right)  \right]  \mathcal{I}_{FF}\right\rangle B_{M^{\prime}}^{+}\nonumber\\
&  +\left\langle \left(  b_{M}^{2}-\bar{b}^{2}\right)  A\right\rangle
B_{M}^{+}=R_{M}^{+}, \label{Bp}%
\end{align}
for the linear combinations of the unknown functions
\begin{equation}
B_{M}^{\pm}=\frac{1}{2}\left(  B_{M}^{\left(  1\right)  }\pm B_{M}^{\left(
2\right)  }\right)  , \label{Bpm}%
\end{equation}
where
\begin{equation}
\Omega=\frac{\omega}{2\Delta},~\eta=\frac{\mathbf{qv}}{2\Delta}, \label{Eta}%
\end{equation}
$b_{M}^{2}\equiv\mathbf{b}_{M}^{\ast}\mathbf{b}_{M}$ and $\mathbf{v}$ is a
vector with the magnitude of the Fermi velocity $\upsilon_{F}$ and the
direction of $\mathbf{n}$. The right-hand sides of Eqs. (\ref{Bm}) and
(\ref{Bp}) are given by%
\begin{align}
R_{M}^{\pm}  &  =\mp\frac{1}{4}\left\langle \operatorname*{Tr}\left[  \left(
\hat{\bm{\sigma}}\mathbf{b}_{M}^{\ast}\right)  \left(  \hat{\tau}%
\hat{\bm{\sigma}}\mathbf{\bar{b}}\mp\hat{\bm{\sigma}}\mathbf{\bar{b}}\hat
{\tau}\right)  \right]  \left(  \Omega+\eta\right)  \mathcal{I}_{FF}%
\right\rangle \nonumber\\
&  \pm\frac{1}{4}\left\langle \operatorname*{Tr}\left[  \left(  \hat
{\bm{\sigma}}\mathbf{b}_{M}^{\ast}\right)  \left(  \hat{\bm{\sigma}}%
\mathbf{\bar{b}}\hat{g}\hat{\tau}^{-}\hat{g}\mp\hat{g}\hat{\tau}^{-}\hat
{g}\hat{\bm{\sigma}}\mathbf{\bar{b}}\right)  \right]  \left(  \Omega
-\eta\right)  \mathcal{I}_{FF}\right\rangle . \label{Fpm}%
\end{align}
The function $A\left(  \mathbf{n},T\right)  $ is of the form
\begin{equation}
A\left(  \mathbf{n}\right)  \equiv\int_{-\infty}^{\infty}d\varepsilon\left(
\frac{1}{2E}\tanh\frac{E}{2T}-\frac{1}{2\varepsilon}\tanh\frac{\varepsilon
}{2T}\right)  , \label{Aex}%
\end{equation}
and the function $\mathcal{I}_{FF}\left(  \omega,\mathbf{q;n},T\right)  $ is
defined in Appendix A. For later use we indicate the explicit form of this
function at $\mathbf{q}\rightarrow0$:
\begin{equation}
\mathcal{I}_{0}\left(  \omega\mathbf{,n}\right)  \equiv\mathcal{I}_{FF}\left(
\omega\mathbf{,q}=0\mathbf{;n}\right)  =\int_{-\infty}^{\infty}\frac
{d\varepsilon}{E}\frac{\Delta^{2}}{4E^{2}-\left(  \omega+i0\right)  ^{2}}%
\tanh\frac{E}{2T}~, \label{FFq0}%
\end{equation}
where
\begin{equation}
E=\sqrt{\varepsilon^{2}+\Delta^{2}\bar{b}^{2}\left(  \mathbf{n}\right)  }~\ .
\label{E}%
\end{equation}

It is necessary to stress that in obtaining Eqs. (\ref{Bp}) and (\ref{Bm}) the
particular form of the gap structure was used, where $\Delta\left(  T\right)
$ is a real scalar and $\mathbf{\bar{b}}\left(  \mathbf{n}\right)  $ is a real
vector. Therefore the present theory does not allow comparison with the theory
of $^{3}$He-A by Wolfe \cite{W}, where $\mathbf{\bar{b}}\left(  \mathbf{n}%
\right)  =\left(  n_{1}+in_{2}\right)  \mathbf{w}$ with $\mathbf{w}$ being a
unit vector in spin space. However, the equations can be easily generalized to
the case of $^{3}$He-B \cite{Wolfe1} where $\mathbf{\bar{b}}\left(
\mathbf{n}\right)  =\mathbf{n}$ (See Appendix B).

\section{Eigenmodes of the order parameter}

We now look for eigensolutions of Eqs.(\ref{Bm}) and (\ref{Bp}) in the BCS
approximation, i.e. regarding the terms $R_{M}^{\pm}$ involving $\hat{\tau
}\equiv\left(  \hat{\tau}_{0},\hat{\bm{\tau}}\right)  $ as sources which they
are in the limit of vanishing Fermi-liquid interactions. We consider the
vertices for density, current, spin-density and spin-current perturbations
interacting with the corresponding vector or axial-vector field $A^{\mu}$.

Expansions (\ref{T1A}) and (\ref{T2A}) imply both the unitary and nonunitary
excited states. The unitary excitations have no spin polarization. In a
nonunitary state the Cooper pairs at point $\mathbf{n}$ have a net average
spin \cite{Leggett75}, \cite{Ketterson}. Evidently the unitary perturbations
can be excited by the vector external fields while the nonunitary
perturbations interact with the axial-vector external fields. In the BCS
approximation the corresponding ordinary vector and axial-vector vertices are
of the form $\hat{\tau}^{\mu}=\left(  1,\mathbf{v}\right)  \hat{1}$ and
$\left(  \mathbf{v\cdot}\hat{\bm{\sigma}},\hat{\bm{\sigma}}\right)  $, respectively.

\subsection{Unitary excitations}

From Eq. (\ref{Fpm}) we obtain for the vector field.
\begin{equation}
R_{M}^{-}=\left(  \frac{\omega}{\Delta}\left\langle \mathbf{\bar{b}b}%
_{M}^{\ast}\mathcal{I}_{FF}\right\rangle ,~\frac{1}{\Delta}\left\langle
\mathbf{v}\left(  \mathbf{qv}\right)  \left(  \mathbf{\bar{b}b}_{M}^{\ast
}\right)  \mathcal{I}_{FF}\right\rangle \right)  ,~R_{M}^{+}=0. \label{Rm}%
\end{equation}
In this case from Eq. (\ref{Bp}) we obtain the trivial solution $B_{M}^{+}=0$,
and inspection of Eq. (\ref{Bm}) reveals that the eigenmodes which can be
excited by the vector external field satisfy the dispersion equation
\begin{align}
&  \det\left\vert 2\left\langle \left[  \left(  \Omega^{2}-\eta^{2}-\bar
{b}^{2}\right)  \mathbf{b}_{M}^{\ast}\mathbf{b}_{M^{\prime}}+\left(
\mathbf{b}_{M}^{\ast}\mathbf{\bar{b}}\right)  \left(  \mathbf{b}_{M^{\prime}%
}\mathbf{\bar{b}}\right)  \right]  \mathcal{I}_{FF}\right\rangle \right.
\nonumber\\
&  \left.  +\delta_{MM^{\prime}}\left\langle \left(  b_{M}^{2}-\bar{b}%
^{2}\right)  A\right\rangle \right\vert =0. \label{detm}%
\end{align}

We examine the eigenmodes with $\mathbf{q}=0$. For the $^{3}P_{2}$ pairing
with $M=0$ one has $\mathbf{\bar{b}=b}_{0}$. Then%
\begin{equation}
\bar{b}^{2}=\frac{1}{2}\left(  1+3n_{3}^{2}\right)  , \label{bsq}%
\end{equation}
and the functions $A\left(  \mathbf{n}\right)  =A\left(  n_{3},y\right)  $ and
$\mathcal{I}_{0}\left(  \omega,\mathbf{n};y\right)  =\mathcal{I}_{0}\left(
\omega,n_{3};y\right)  $ are axially symmetric at $\mathbf{q}=0$. Here and
below we denote $y=\Delta\left(  T\right)  /T$. In the case $\mathbf{q}=0$ the
azimuth-angle integrals can be performed in Eq. (\ref{detm}) making use of the
orthogonality relations
\begin{equation}
\int\frac{d\varphi}{2\pi}\mathbf{b}_{M}^{\ast}\mathbf{b}_{M^{\prime}}%
=\delta_{MM^{\prime}}b_{M}^{2}, \label{bbdfi}%
\end{equation}%
\begin{equation}
\int\frac{d\varphi}{2\pi}\left(  \mathbf{b}_{M}^{\ast}\mathbf{b}_{0}\right)
\left(  \mathbf{b}_{M^{\prime}}\mathbf{b}_{0}\right)  =\delta_{M,M^{\prime}%
}\left(  \mathbf{b}_{M}^{\ast}\mathbf{b}_{0}\right)  \left(  \mathbf{b}%
_{M}\mathbf{b}_{0}\right)  , \label{bbdfi1}%
\end{equation}
which can be easily verified. As a result after some algebraic manipulations,
the dispersion equation (\ref{detm}) at $\mathbf{q}=0$ can be obtained in the
form%
\begin{equation}
\prod\limits_{M}f_{M}\left(  \omega,y\right)  =0 \label{detv}%
\end{equation}
with
\begin{equation}
f_{M}\left(  \omega,y\right)  =\left\langle \left(  b_{M}^{2}-\bar{b}%
^{2}\right)  A+2\left[  \left(  \frac{\omega^{2}}{4\Delta^{2}}-\bar{b}%
^{2}\right)  b_{M}^{2}+\left(  \mathbf{b}_{M}^{\ast}\mathbf{\bar{b}}\right)
\left(  \mathbf{b}_{M}\mathbf{\bar{b}}\right)  \right]  \mathcal{I}%
_{0}\right\rangle . \label{fM}%
\end{equation}
We thus obtain separate eigenvalue equations for each value of $M=0,\pm1,\pm2$
in the vector channel:
\begin{equation}
\left\langle b_{M}^{2}\mathcal{I}_{0}\right\rangle \frac{\omega^{2}}%
{2\Delta^{2}}=2\left\langle \left[  \bar{b}^{2}b_{M}^{2}-\left(
\mathbf{b}_{M}^{\ast}\mathbf{\bar{b}}\right)  \left(  \mathbf{b}%
_{M}\mathbf{\bar{b}}\right)  \right]  \mathcal{I}_{0}\right\rangle
-\left\langle \left(  b_{M}^{2}-\bar{b}^{2}\right)  A\right\rangle .
\label{wEq}%
\end{equation}
At $\omega>2\Delta\bar{b}$ the function $\mathcal{I}_{0}\left(  \omega
,n_{3};y\right)  $ has the imaginary part owing to the breaking and formation
of Cooper pairs. Therefore the undamped collective modes could exist only at
$0\leq\omega<2\Delta\sqrt{1/2}$.

From Eq. (\ref{wEq}) we obtain the eigenvalue equation for a perturbation with
$M=0$,
\begin{equation}
\omega_{0}^{2}\int_{0}^{1}dn_{3}\left(  1+3n_{3}^{2}\right)  \mathcal{I}%
_{0}\left(  \omega_{0},n_{3};y\right)  =0, \label{V0eq}%
\end{equation}
yielding%
\begin{equation}
\omega_{0}\left(  y\right)  \equiv0. \label{V0}%
\end{equation}

At $M=\pm1$ we obtain the equation
\begin{align}
&  \int_{0}^{1}dn_{3}\left[  \frac{1}{2\Delta^{2}}\omega_{1}^{2}\left(
1+n_{3}^{2}\right)  -\left(  3n_{3}^{2}+4n_{3}^{4}+1\right)  \right]
\mathcal{I}_{0}\left(  \omega_{1},n_{3};y\right) \nonumber\\
&  =-\frac{1}{3}\int_{0}^{1}dn_{3}\left(  1-3n_{3}^{2}\right)  A\left(
n_{3};y\right)  , \label{V1}%
\end{align}
which has no real solutions in the domain $0\leq\omega<2\Delta\sqrt{1/2}$.

For $M=\pm2$ we find $\omega\left(  \mathbf{q}=0\right)  =\omega_{2}\left(
y\right)  $ which satisfy the dispersion equation
\begin{align}
&  \int_{0}^{1}dn_{3}\left[  \frac{3}{4\Delta^{2}}\omega_{2}^{2}\left(
1-n_{3}^{2}\right)  -\frac{3}{4}\left(  1-n_{3}^{2}\right)  \left(
1+7n_{3}^{2}\right)  \right]  \mathcal{I}_{0}\left(  \omega_{2},n_{3};y\right)
\nonumber\\
&  =-\int_{0}^{1}dn_{3}\left(  1-3n_{3}^{2}\right)  A\left(  n_{3};y\right)  .
\label{V2}%
\end{align}

\subsection{Nonunitary excitations}

For the axial-vector field from Eq. (\ref{Fpm}) we find
\begin{equation}
R_{M}^{-}=0,~R_{M}^{+}=-i\left(  \frac{1}{\Delta}\left\langle \left(
\mathbf{qv}\right)  \left(  \mathbf{b}_{M}^{\ast}\mathbf{\times\bar{b}%
}\right)  \mathbf{v}\mathcal{I}_{FF}\right\rangle ,~\frac{\omega}{\Delta
}\left\langle \left(  \mathbf{b}_{M}^{\ast}\mathbf{\times\bar{b}}\right)
\mathcal{I}_{FF}\right\rangle \right)  . \label{Rp}%
\end{equation}
In this case in Eq. (\ref{Bm}) the right-hand side vanishes and we obtain the
trivial solution $B_{M}^{-}=0$. Inspection of Eqs. (\ref{Bp}) reveals that the
eigenmodes of the pseudo-vector current satisfy the dispersion equation%
\begin{align}
&  \det\left\vert 2\left\langle \left[  \left(  \Omega^{2}-\eta^{2}\right)
\mathbf{b}_{M}^{\ast}\mathbf{b}_{M^{\prime}}-\left(  \mathbf{b}_{M}^{\ast
}\mathbf{\bar{b}}\right)  \left(  \mathbf{b}_{M^{\prime}}\mathbf{\bar{b}%
}\right)  \right]  \mathcal{I}_{FF}\right\rangle \right. \nonumber\\
&  \left.  +\delta_{MM^{\prime}}\left\langle \left(  b_{M}^{2}-\bar{b}%
^{2}\right)  A\right\rangle \right\vert =0. \label{det}%
\end{align}
In the limit $q=0$ the manipulations with making use of Eqs. (\ref{bbdfi}) and
(\ref{bbdfi1}) allow one to obtain Eq. (\ref{det}) in the form%
\begin{equation}
\prod\limits_{M}\tilde{f}_{M}\left(  \omega,y\right)  =0, \label{deta}%
\end{equation}
where
\begin{equation}
\tilde{f}_{M}\left(  \omega,y\right)  =\left\langle \left(  b_{M}^{2}-\bar
{b}^{2}\right)  A+2\left[  \frac{\omega^{2}}{4\Delta^{2}}b_{M}^{2}-\left(
\mathbf{b}_{M}^{\ast}\mathbf{\bar{b}}\right)  \left(  \mathbf{b}%
_{M}\mathbf{\bar{b}}\right)  \right]  \mathcal{I}_{0}\right\rangle .
\label{fMt}%
\end{equation}
The eigenvalues $\tilde{\omega}_{M}$ for $M=0,\pm1,\pm2$ in the axial-vector
channel can be found from the dispersion equation%
\begin{equation}
\left\langle \left\vert b_{M}\right\vert ^{2}\mathcal{I}_{0}\right\rangle
\frac{\omega^{2}}{2\Delta^{2}}=2\left\langle \left(  \mathbf{b}_{M}^{\ast
}\mathbf{\bar{b}}\right)  \left(  \mathbf{b}_{M}\mathbf{\bar{b}}\right)
\mathcal{I}_{0}\right\rangle -\left\langle \left(  \left\vert b_{M}\right\vert
^{2}-\bar{b}^{2}\right)  A\right\rangle . \label{detA}%
\end{equation}

For $M=0$ we obtain%
\begin{equation}
\int_{0}^{1}dn_{3}\left[  \frac{1}{2\Delta^{2}}\omega^{2}\left(  1+3n_{3}%
^{2}\right)  -\left(  1+3n_{3}^{2}\right)  ^{2}\right]  \mathcal{I}_{0}\left(
\omega,n_{3};y\right)  =0. \label{A3}%
\end{equation}
This equation has no real solutions at $\left\vert \omega\right\vert
<2\Delta\sqrt{1/2}$.

For $M=\pm1$ Eq. (\ref{detA}) yields $\omega\left(  \mathbf{q}=0\right)
=\tilde{\omega}_{1}\left(  y\right)  $ satisfying%
\begin{align}
&  \int_{0}^{1}dn_{3}\left[  \frac{1}{2\Delta^{2}}\tilde{\omega}_{1}%
^{2}\left(  1+n_{3}^{2}\right)  -n_{3}^{2}\left(  1-n_{3}^{2}\right)  \right]
\mathcal{I}_{0}\left(  \tilde{\omega}_{1},n_{3};y\right) \nonumber\\
&  =-\frac{1}{3}\int_{0}^{1}dn_{3}\left(  1-3n_{3}^{2}\right)  A\left(
n_{3};y\right)  . \label{A1}%
\end{align}
At $M=\pm2$ we find $\omega\left(  \mathbf{q}=0\right)  =\tilde{\omega}%
_{2}\left(  y\right)  $ satisfying the dispersion equation%
\begin{align}
&  \int_{0}^{1}dn_{3}\left[  \frac{1}{\Delta^{2}}\tilde{\omega}_{2}^{2}\left(
1-n_{3}^{2}\right)  -\left(  1-n_{3}^{2}\right)  ^{2}\right]  \mathcal{I}%
_{0}\left(  \tilde{\omega}_{2},n_{3};y\right) \nonumber\\
&  =-\frac{4}{3}\int_{0}^{1}dn_{3}\left(  1-3n_{3}^{2}\right)  A\left(
n_{3};y\right)  . \label{A2}%
\end{align}

\begin{figure}[h]
\includegraphics{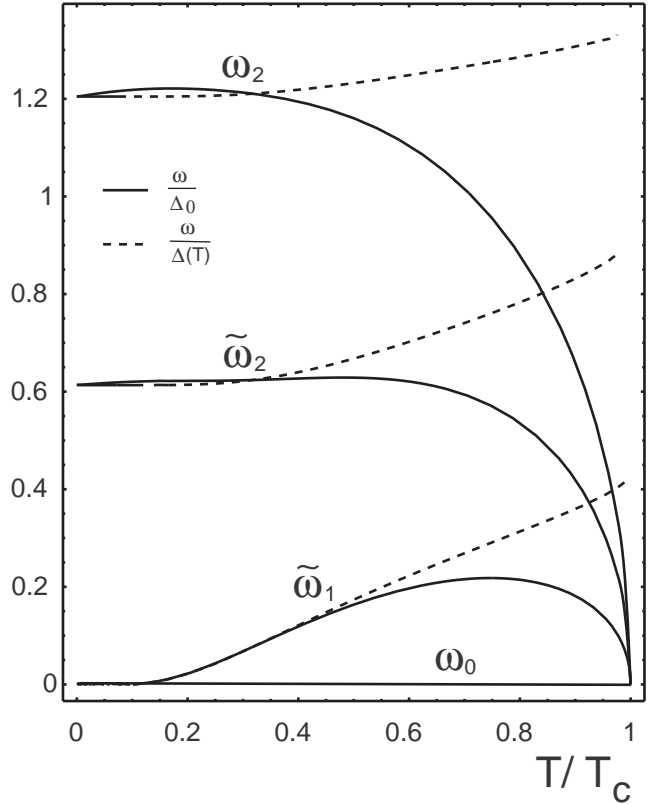}\caption{Collective frequencies of the neutron
$^{3}P_{2}$ condensate at $q=0$ in units of $\Delta_{0}\equiv\Delta\left(
T=0\right)  $ and in units of $\Delta\left(  T\right)  $ versus reduced
temperature in the BCS approximation. }%
\label{fig2}%
\end{figure}

The eigen frequencies of the collective excitations in the limit
$q\rightarrow0$ are represented in Fig. \ref{fig2} versus reduced temperature
$T/T_{c}$. In this plot, the oscillation frequencies are shown both in units
of $\Delta_{0}\equiv\Delta\left(  T=0\right)  =1.\,\allowbreak68T_{c}$ (See
Appendix C) and in units of $\Delta\left(  T\right)  $. We assume \cite{YKL}
$y\left(  T/Tc\right)  =\sqrt{2}\sqrt{1-T/T_{c}}(0.7893+1.188T_{c}/T)$.

\section{Discussion and conclusion}

Since the equilibrium state of the condensate corresponds to $M=0$ the
variables $B_{0}^{\left(  1,2\right)  }$ are associated with the broken gauge
symmetry. In Fig. \ref{fig2}, the lowest curve $\omega_{0}=0$ represents the
corresponding Goldstone's mode $\omega=\omega_{0}\left(  \mathbf{q}\right)  $
at $q=0$. The natural appearance of this mode is caused by spontaneous
breaking of the baryon number owing to the Cooper condensation. The collective
motion of the condensate in this wave is a periodic variation of the total
phase without a change of the order parameter structure. The previous analysis
\cite{L11a} has shown that, at $q>0$, the Goldstone's mode is anisotropic and
strongly renormalized by the Fermi-liquid interactions. The sound velocity is
found strongly dependent on the Landau parameters describing the residual
particle-hole interactions.

It was expected \cite{Bed} that the spontaneous breaking of rotational
invariance owing to the $^{3}P_{2}$ pairing leads to the appearance of three
more Goldstone bosons (angulons). We found no Goldstone modes associated with
the breaking of rotational symmetry. In our model, the variables $B_{\pm
1}^{\left(  1,2\right)  }$ are associated with a flapping motion of the total
angular momentum which represents the nonunitary state, where the excited
Cooper pairs have a nonzero average spin. In other words this can be imagined
as a departure of the symmetry axis of the bound pair about the symmetry axis
of the equilibrated condensate. This is equivalent to oscillations of the
preferred direction of the Cooper pair about the axis of the energy gap in the
quasiparticle energy. The quasiparticle readjustment can not follow this rapid
motion. As a result instead of the angulons we found the modes $\tilde{\omega
}_{1}$ which correspond to $M=\pm1$ and look like the "normal-flapping" mode
in $^{3}$He-A \cite{W}. The temperature variation of $\tilde{\omega}_{1}$ is
naturally explained: the moment of inertia associated with the quasiparticles
declines owing to a decrease of number of thermal quasiparticles along with a
lowering of the temperature. At zero temperature the restoring force is
expected to vanish along with the number of thermal quasiparticles.
Accordingly, the frequency of these modes passes through a maximum and tends
to zero for $T\rightarrow0$.

The variables $B_{\pm2}^{\left(  1,2\right)  }$ are found as $q\rightarrow0$
to oscillate at frequencies $\omega=\omega_{2}\left(  y\right)  $ and
$\omega=\tilde{\omega}_{2}\left(  y\right)  $ with $\omega_{2}\left(
T=0\right)  =1.20\Delta\left(  0\right)  $ and $\tilde{\omega}_{2}\left(
T=0\right)  =0.61\Delta\left(  0\right)  $. To understand the nature of these
excitations we introduce three spontaneous orbital axes of the gap
$\mathbf{j},\mathbf{k},\mathbf{l}$ with $\mathbf{j}\perp\mathbf{k}$ and
$\mathbf{l}=\mathbf{j\times k}$ being the symmetry axis. Then the equilibrium
gap matrix of the $^{3}P_{2}$ condensate with $M=0$ at the point $\mathbf{n}$
on the Fermi surface can be written as
\[
\hat{D}=\frac{1}{\sqrt{2}}\left(
\begin{array}
[c]{cc}%
\left(  \mathbf{j}-i\mathbf{k}\right)  \mathbf{n} & 2\mathbf{ln}\\
2\mathbf{ln} & -\left(  \mathbf{j}+i\mathbf{k}\right)  \mathbf{n}%
\end{array}
\right)  .
\]
Variation of this matrix caused by the modes with $M=\pm2$ is of the form (we
omit the superscript in $B_{2}^{1}$)
\[
\delta\hat{D}_{\left\vert 2\right\vert }=e^{i\omega t}\left(
\begin{array}
[c]{cc}%
-B_{2}\left(  \mathbf{j}+i\mathbf{k}\right)  \mathbf{n} & 0\\
0 & B_{-2}\left(  \mathbf{j}-i\mathbf{k}\right)  \mathbf{n}%
\end{array}
\right)  .
\]

This form indicates that the oscillations occur in the plane $(\mathbf{j}%
,\mathbf{k})$. The admixture of the $\delta\hat{D}_{2}$ to $\hat{D}$ can be
imagined as a clapping motion of $\mathbf{j}$ and $\mathbf{k}$ about their
equilibrium relative angle of $\pi/2$.

According to Eqs. (\ref{Rm}), in the vector channel of excitation one has
$B_{-2}=B_{2}^{\ast}$, and the excited state $\mathfrak{D}=\hat{D}+\delta
\hat{D}_{\left\vert 2\right\vert }$ is unitary. This means the $\omega_{2}$
excitations correspond to the states with no net spin polarization. In the
axial-vector channel, from Eqs. (\ref{Rp}) one obtains $B_{-2}=-B_{2}^{\ast}$,
and the excited state $\mathfrak{D}=\hat{D}+\delta\hat{D}_{\left\vert
2\right\vert }$ is nonunitary, i.e. for $\tilde{\omega}_{2}$ excitations the
average spin expectation value is nonzero. (The same takes place for the
flapping mode $\tilde{\omega}_{1}$.) This allows one to term these
oscillations by spin waves.

The flapping mode $\tilde{\omega}_{1}$ needs special discussion. The
oscillation frequency for this mode is small, $\tilde{\omega}_{1}^{2}%
\ll4\Delta^{2}$. This allows one to neglect $\omega^{2}$ in the integrand of
Eq. (\ref{FFq0}) thus obtaining%
\begin{equation}
\mathcal{I}_{0}\left(  \tilde{\omega}_{1}\mathbf{,n}\right)  \simeq\tilde
{I}_{0}\left(  \mathbf{n}\right)  \equiv\frac{\Delta^{2}}{4}\int_{-\infty
}^{\infty}\frac{d\varepsilon}{E^{3}}\tanh\frac{E}{2T}\text{ } \label{It}%
\end{equation}
with $E=\sqrt{\varepsilon^{2}+\Delta^{2}\bar{b}^{2}\left(  n_{3}\right)  }$.

After this simplification we find \ the analytic solution of the form%
\begin{equation}
\tilde{\omega}_{1}^{2}=\Delta^{2}\frac{2\int_{0}^{1}dn_{3}\left[  3n_{3}%
^{2}\left(  1-n_{3}^{2}\right)  \tilde{I}_{0}-\left(  1-3n_{3}^{2}\right)
A\right]  }{3\int_{0}^{1}dn_{3}\left(  1+n_{3}^{2}\right)  \tilde{I}_{0}}.
\label{w1}%
\end{equation}
In previous works \cite{L10c} the right-hand side of Eq. (\ref{w1}) was
evaluated in the average-angle approximation assuming that the anisotropic gap
in the quasiparticle energy is replaced by its average-angle magnitude,
$\bar{b}^{2}\Delta^{2}\left(  T\right)  \rightarrow\Delta^{2}\left(  T\right)
$. Such approach results in $\tilde{\omega}_{1}^{\mathsf{av}}=\Delta\left(
T\right)  /\sqrt{5}$ with a simple temperature dependence of the excitation
energy only through the gap amplitude. Owing to the gap anisotropy the energy
of the spin density oscillations depends more dramatically on the temperature,
as shown in Fig. \ref{fig2}. As can be seen the gap anisotropy leads to a
strong decreasing of the level energy along with a lowering of the
temperature. This is to suppress substantially the neutrino energy losses
caused by the spin wave decays because the rate of neutrino losses is strongly
dependent on the wave energy. Since we have found also the new collective
modes, which are kinematically able to decay into neutrino pairs, the problem
of neutrino emissivity of the $^{3}P_{2}$ condensate at lowest temperatures is
to be revisited.

It is necessary to note finally that the Fermi-liquid effects can somewhat
modify the collective frequencies as compared to that found in the BCS approximation.

\setcounter{equation}{0}
\begin{appendix}

\section{Functions used}

\label{sec:A} We denote as $L_{X,X}\left(  \omega,\mathbf{q;p}\right)  $ the
analytical continuation of the Matsubara sums:%
\begin{equation}
L_{XX^{\prime}}\left(  \omega_{\eta},\mathbf{p+}\frac{\mathbf{q}}%
{2}\mathbf{;p-}\frac{\mathbf{q}}{2}\right)  =T\sum_{\kappa}X\left(
P_{+}\right)  X^{\prime}\left(  P_{-}\right)  ,\label{LXX}%
\end{equation}
where $X,X^{\prime}\in G,F,G^{-}$ and operate with integrals over the
quasiparticle energy:%
\begin{equation}
\mathcal{I}_{XX^{\prime}}\left(  \omega,\mathbf{n,q};T\right)  \equiv
\int_{-\infty}^{\infty}d\varepsilon_{p}L_{XX^{\prime}}\left(  \omega
,\mathbf{p+}\frac{\mathbf{q}}{2}\mathbf{,p-}\frac{\mathbf{q}}{2}\right)
.\label{loop}%
\end{equation}
These are functions of $\omega$, $\mathbf{q}$ and the direction of a
quasiparticle momentum $\mathbf{n}$, except the function $\mathcal{I}%
_{G_{N}G_{N}^{-}}$ which is to be evaluated for $\omega=0$, $\mathbf{q}=0$ in
all cases. One can easily check that%
\begin{equation}
\mathcal{I}_{G^{-}G}=\mathcal{I}_{GG^{-}}~,~\mathcal{I}_{GF}=-\mathcal{I}%
_{FG}~,~\mathcal{I}_{G^{-}F}=-\mathcal{I}_{FG^{-}}~.\label{Leg}%
\end{equation}
We use also the following relations valid for arbitrary $\omega,\mathbf{q},T$
\begin{equation}
\mathcal{I}_{G^{-}F}=\frac{\omega-\mathbf{qv}}{2\Delta}\mathcal{I}%
_{FF}~,\label{gf}%
\end{equation}%
\begin{equation}
\mathcal{I}_{FG}=\frac{\omega+\mathbf{qv}}{2\Delta}\mathcal{I}_{FF}%
~,\label{ff1}%
\end{equation}
and%
\begin{equation}
\mathcal{I}_{GG^{-}}-\mathcal{I}_{G_{N}G_{N}^{-}}+\bar{b}^{2}\mathcal{I}%
_{FF}=A+\frac{\omega^{2}-\left(  \mathbf{qv}\right)  ^{2}}{2\Delta^{2}%
}\mathcal{I}_{FF},\label{FFp}%
\end{equation}
The function $A\left(  \mathbf{n}\right)  $ is defined by Eq. (\ref{A}) and
can be written as%
\begin{equation}
A\left(  \mathbf{n}\right)  =\left[  \mathcal{I}_{GG^{-}}-\mathcal{I}%
_{G_{N}G_{N}^{-}}+\bar{b}^{2}\mathcal{I}_{FF}\right]  _{\omega=0,\mathbf{q}%
=0}.\label{Ar}%
\end{equation}
The explicit form of this function is given in Eq. (\ref{Aex}).

\section{Eigenmodes in $^{3}$He-B}

\label{sec:B} In the liquid $^{3}$He-B, the pairing occurs owing to the
exchange central interaction which is independent of the total angular
momentum of the bound pair, $J$. In this case the interaction in the channel
$S=1$, $L=1$ can be written in the same form as given in Eq. (\ref{Gam}) but
the summation over the total angular momentum $J=0,1,2$ is to be added with
$-J\leq M\leq J$. Then instead of Eqs. (\ref{detm}) and (\ref{det}) one
obtains, respectively,%
\begin{align}
&  \det\left\vert ~\left\langle \left(  \mathbf{b}_{J^{\prime}M^{\prime}%
}\mathbf{b}_{JM}^{\ast}-\delta_{JJ^{\prime}}\delta_{MM^{\prime}}%
\mathbf{\bar{b}}^{2}\right)  A\right\rangle \right.  \nonumber\\
&  \left.  +\left\langle \left[  \left(  \frac{\omega^{2}-\left(
\mathbf{qv}\right)  ^{2}}{2\Delta^{2}}-2\bar{b}^{2}\right)  \left(
\mathbf{b}_{J^{\prime}M^{\prime}}\mathbf{b}_{JM}^{\ast}\right)  +2\left(
\mathbf{b}_{JM}^{\ast}\mathbf{\bar{b}}\right)  \left(  \mathbf{b}_{J^{\prime
}M^{\prime}}\mathbf{\bar{b}}\right)  \right]  \mathcal{I}_{0}\right\rangle
\right\vert =0,\label{detVHe}%
\end{align}
and%
\begin{align}
&  \det\left\vert ~\left\langle \left(  \mathbf{b}_{J^{\prime}M^{\prime}%
}\mathbf{b}_{JM}^{\ast}-\delta_{JJ^{\prime}}\delta_{MM^{\prime}}%
\mathbf{\bar{b}}^{2}\right)  A\right\rangle \right.  \nonumber\\
&  \left.  +\left\langle \left[  \frac{\omega^{2}-\left(  \mathbf{qv}\right)
^{2}}{2\Delta^{2}}\left(  \mathbf{b}_{J^{\prime}M^{\prime}}\mathbf{b}%
_{JM}^{\ast}\right)  -2\left(  \mathbf{b}_{JM}^{\ast}\mathbf{\bar{b}}\right)
\left(  \mathbf{b}_{J^{\prime}M^{\prime}}\mathbf{\bar{b}}\right)  \right]
\mathcal{I}_{0}\right\rangle \right\vert =0,\label{detpHe}%
\end{align}
where the vectors $\mathbf{b}_{JM}$ for $J=2$ are given by Eq. (\ref{b012})
and for $J=0,1$ one has
\begin{align}
\mathbf{b}_{0,0} &  =\left(  -n_{1},-n_{2},-n_{3}\right)  ,\nonumber\\
\mathbf{b}_{1,0} &  =\sqrt{\frac{3}{2}}\left(  in_{2},-in_{1},0\right)
,\nonumber\\
\mathbf{b}_{1,1} &  =\mathbf{b}_{1,-1}^{\ast}=\sqrt{\frac{3}{4}}\left(
-n_{3},-in_{3},n_{1}+in_{2}\right)  ,\label{J1}%
\end{align}
In $^{3}$He-B, the equilibrium state of the condensate corresponds to the
bound pairs with $J=0$. The order parameter in such a system is $\mathbf{\bar
{b}}=\mathbf{b}_{0,0}$, and the energy gap $\Delta\bar{b}\left(
\mathbf{n}\right)  =\Delta$ is isotropic ($\bar{b}^{2}=1$). The latter means
that, in Eqs. (\ref{detVHe}), (\ref{detpHe}), the functions $A$ and
$\mathcal{I}_{0}$ are isotropic and can be moved beyond the integrals. Using
also the orthogonality condition, $\left\langle \mathbf{b}_{JM}\mathbf{b}%
_{J^{\prime}M^{\prime}}^{\ast}\right\rangle =\delta_{JJ^{\prime}}%
\delta_{MM^{\prime}}$, we find
\begin{equation}
~\left\langle \mathbf{b}_{J^{\prime}M^{\prime}}\mathbf{b}_{JM}^{\ast}%
-\delta_{JJ^{\prime}}\delta_{MM^{\prime}}\mathbf{\bar{b}}^{2}\right\rangle
=0~,\label{A0}%
\end{equation}
thus obtaining from Eq. (\ref{detVHe}) the dispersion equation
\begin{equation}
\det\left\vert \left\langle \left(  \frac{\omega^{2}-\left(  \mathbf{qv}%
\right)  ^{2}}{2\Delta^{2}}-2\right)  \left(  \mathbf{b}_{J^{\prime}M^{\prime
}}\mathbf{b}_{JM}^{\ast}\right)  +2\left(  \mathbf{b}_{JM}^{\ast}%
\mathbf{\bar{b}}\right)  \left(  \mathbf{b}_{J^{\prime}M^{\prime}}%
\mathbf{\bar{b}}\right)  \right\rangle \right\vert =0.\label{dEqBp}%
\end{equation}
We examine solutions to this equation for $\mathbf{q}=0$. A straightforward
calculation of the angle integrals results in%
\begin{equation}
\left\langle \left(  \mathbf{b}_{JM}^{\ast}\mathbf{\bar{b}}\right)  \left(
\mathbf{b}_{J^{\prime}M^{\prime}}\mathbf{\bar{b}}\right)  \right\rangle
=\delta_{JJ^{\prime}}\delta_{MM^{\prime}}\left(  \delta_{J0}\delta_{M0}%
+\frac{2}{5}\delta_{J2}\right)  .\label{bbbb}%
\end{equation}
Implying $J=0,1,2$ and $\left\vert M\right\vert \leq J$ we obtain the
dispersion equation
\begin{equation}
\Omega^{2}\left(  \Omega^{2}-1\right)  ^{3}\left(  \Omega^{2}-\frac{3}%
{5}\right)  ^{5}=0,\label{vmodes}%
\end{equation}
where we denote
\begin{equation}
\Omega=\frac{\omega}{2\Delta\left(  T\right)  }.\label{Omega}%
\end{equation}
From this equation we find one mode with frequency $\omega=0$, five degenerate
modes with frequency $\omega=\sqrt{12/5}\Delta$, and three degenerate modes
with frequency $\omega=2\Delta$. In the same way starting from Eq.
(\ref{detpHe}) we obtain the dispersion equation for eigenmodes of spin
density and spin-current density which can be excited by the axial-vector
external field%
\begin{equation}
\left(  \Omega^{2}-1\right)  \left(  \Omega^{2}\right)  ^{3}\left(  \Omega
^{2}-\frac{2}{5}\right)  ^{5}=0.\label{amodes}%
\end{equation}
From this equation we find one mode with frequency $\omega=2\Delta$, five
degenerate modes with frequency $\omega=\sqrt{8/5}\Delta$, and three
degenerate modes with frequency $\omega=0$. This completes the enumeration of
the 18 B-phase collective modes for $\mathbf{q}=0$. Note these results are
identical to ones obtained earlier in Refs. \cite{Wolfe1}, \cite{Ketterson}.

\section{Critical temperature}

\label{sec:C} Combining Eq. (\ref{GAP}) for $T>0$ with the same equation for
$T=0$ one can obtain the equality%
\begin{equation}
\left\langle \bar{b}^{2}\int_{-\infty}^{\infty}d\varepsilon\left[  \left(
\varepsilon^{2}+\Delta_{0}^{2}\bar{b}^{2}\right)  ^{-1/2}-\left(
\varepsilon^{2}+\Delta^{2}\bar{b}^{2}\right)  ^{-1/2}\right]  \right\rangle
=-\left\langle 2\bar{b}^{2}\int_{-\infty}^{\infty}\frac{d\varepsilon}%
{E}\left(  e^{E/T}+1\right)  ^{-1}\right\rangle ,\label{T0T}%
\end{equation}
where $E=\sqrt{\varepsilon^{2}+\Delta^{2}\bar{b}^{2}}$. Performing the
integral in the left-hand side we find%
\begin{equation}
\ln\frac{\Delta_{0}}{\Delta}=2\left\langle \bar{b}^{2}\int_{0}^{\infty}%
\frac{dx}{\sqrt{x^{2}+y^{2}\bar{b}^{2}}}\left(  e^{\sqrt{x^{2}+y^{2}\bar
{b}^{2}}}+1\right)  ^{-1}\right\rangle .\label{lgint}%
\end{equation}
The internal integral in the right-hand side is calculated in Ref. \cite{lif}.
Close to the transition point $T\rightarrow T_{c}$ one obtains%
\begin{equation}
\ln\frac{\Delta_{0}}{\Delta}=\left\langle \bar{b}^{2}\left(  \ln\frac{\pi
T}{\gamma\Delta\bar{b}}+\frac{7\zeta\left(  3\right)  }{8\pi^{2}}\frac
{\Delta^{2}}{T^{2}}\right)  \right\rangle ,\label{lg}%
\end{equation}
where $\ln\gamma=C=0.577$ is Euler's constant. According to this equation the
gap vanishes at the temperature%
\begin{equation}
T_{c}=\Delta_{0}\frac{\gamma}{\pi}\exp\left\langle \bar{b}^{2}\ln\bar
{b}\right\rangle .\label{Tc}%
\end{equation}
We evaluate this relation for $\bar{b}^{2}=\left(  1+3n_{3}^{2}\right)  /2$ to
find
\begin{equation}
\frac{T_{c}}{\Delta_{0}}=\frac{\sqrt{2}\gamma}{\pi}e^{\frac{1}{27}\pi\sqrt
{3}-\frac{1}{2}}=0.595.\label{TcD}%
\end{equation}
\end{appendix}

\end{document}